\documentclass[prb,twocolumn,superscriptaddress,floats,showpacs]{revtex4}

 \usepackage[dvips]{graphicx}
 \usepackage[dvips]{graphics}

\usepackage[dvips]{graphicx}

\begin{document}
\title{Evidence for competing magnetic instabilities in underdoped $\rm YBa_2Cu_3O_{6+x}$}

\author{V. Bal\'edent}
\affiliation{Laboratoire L\'eon Brillouin, CEA-CNRS, CE-Saclay, 91191 Gif sur Yvette, France}

\author{D. Haug}
\affiliation{Max-Planck-Institut f\"ur Festk\"orperforschung, 70569 Stuttgart, Germany}

\author{Y. Sidis}
\affiliation{Laboratoire L\'eon Brillouin, CEA-CNRS, CE-Saclay, 91191 Gif sur Yvette, France}

\author{V. Hinkov}
\affiliation{Max-Planck-Institut f\"ur Festk\"orperforschung, 70569 Stuttgart, Germany} \affiliation{Department of Physics and Astronomy, University of British Columbia, Vancouver, Canada V6T 1Z1}

\author{C. T. Lin}
\affiliation{Max-Planck-Institut f\"ur Festk\"orperforschung, 70569 Stuttgart, Germany}

\author{P. Bourges}
\email{philippe.bourges@cea.fr} \affiliation{Laboratoire L\'eon Brillouin, CEA-CNRS, CE-Saclay, 91191 Gif sur Yvette, France}

\date{\today}


\begin{abstract}
 We report a polarized neutron scattering study of the orbital-like magnetic order in strongly underdoped 
${\rm YBa_2Cu_3O_{6.45}}$ and ${\rm YBa_2(Cu_{0.98}Zn_{0.02})_3O_{6.6}}$. Their hole doping levels are located on both sides of the critical doping $p_{MI}$ of a metal-insulator transition inferred from transport measurements. Our study reveals a drop down of the orbital-like order slightly below  $p_{MI}$ with a steep decrease  of both the ordering temperature $T_{mag}$ and the ordered moment. Above $p_{MI}$, substitution of quantum impurities does not change $T_{mag}$, whereas it lowers significantly the bulk ordered moment. The modifications of the orbital-like magnetic order are interpreted in terms of a competition with electronic liquid crystal phases around $p_{MI}$. This competition gives rise to a mixed magnetic state in ${\rm YBa_2Cu_3O_{6.45}}$ and a phase separation in ${\rm YBa_2(Cu_{0.98}Zn_{0.02})_3O_{6.6}}$. 
\end{abstract}

\maketitle


\section{\label{Intro} Introduction}

\begin{figure}[t]
\includegraphics[width=8cm,angle=0]{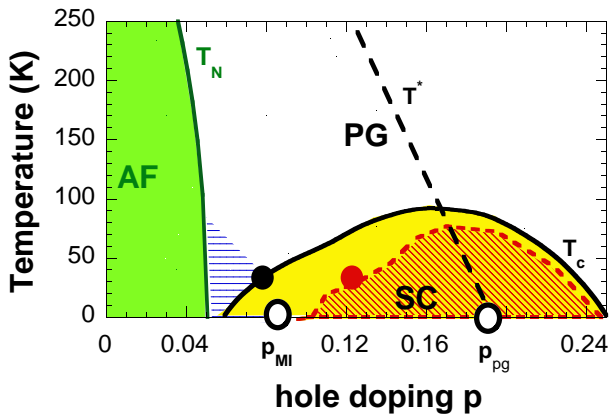}
\caption {
(color online) Schematic phase diagram of ${\rm YBa_2Cu_3O_{6+x}}$ as a function of hole doping $p$; At low doping, the system is a Mott insulator and becomes an antiferromagnet (AF) below the N\'eel temperature $T_N$ (green area) \cite{Casalta,Coneri}. Upon increasing hole doping, it enters the superconducting (SC) state below the critical temperature $T_c$ (black bold line). $p$ is deduced from ref.~\cite{Doping-YBCO}. The hatched superconducting area illustrates the reduction of the superconducting critical temperature $T_c$ in ${\rm YBa_2(Cu_{0.98}Zn_{0.02})_3O_{6+x}}$. The hole doping of both samples, that are studied here, are indicated in the phase diagram by a black point for Y645 ($p$=0.08, $T_c$=35 K) and a red point for Y66-Zn ($p$=0.12, $T_c$=30 K).
}
\label{Fig-intro}
\end{figure}

There has been a long-standing debate among condensed-matter physicists about the origin of the pseudo-gap (PG) state in high-temperature superconducting (SC) cuprates \cite{Norman}. Two theoretical approaches have been opposed: in the former, the pseudogap state is a precursor of the SC state, in the later, its corresponds to another state of matter, competing with superconductivity. Supporting the second scenario, polarized neutron diffraction has recently revealed the existence of a 3D long range magnetic phase, hidden inside the pseudo-gap state of underdoped ${\rm YBa_2Cu_3O_{6+x}}$ \cite{Fauque,Mook}. This phase is likely to be a generic feature of the pseudo-gap state, as it has been later observed in another cuprate family ${\rm HgBa_2CuO_{4+\delta}}$ \cite{Li}. In both cuprates families, its ordering temperature $T_{mag}$ decreases linearly as a function of hole doping ($p$) and vanishes near $p_{pg}$$\sim$0.19 (Fig.~\ref{Fig-intro}), the end point of the pseudo-gap phase according to thermodynamic measurements \cite{Tallon}. $T_{mag}$ matches the pseudo-gap temperature $T^{\star}$  deduced from resistivity measurements \cite{Li,Ito} or appears slightly lower than $T^{\star}$ in other resistivity \cite{Alloul-2010} and NMR Knight shift studies \cite{Alloul-1989} depending of the criterion to determine $T^{\star}$. Thus, $T^{\star}$  could be viewed as the onset temperature associated with the magnetic ordering. In addition, a singular point occuring at $T_{mag}$ was identified in susceptibility measurements carried out in underdoped 
${\rm YBa_2Cu_3O_{6+x}}$ and interpreted as a thermodynamic indication for the existence of a phase transition in the pseudogap state \cite{Leridon}.

The magnetic phase can be described as a {\bf Q}=0 antiferromagnetic ({\bf Q}=0 AF) state:  time reversal symmetry is broken, but lattice translation invariance is preserved, since the same AF pattern develops within each unit cell. The occurence of such a magnetic state has been  predicted in the circulating current (CC) theory of the pseudo-gap proposed by C.M.Varma \cite{Varma}. In this theory, staggered current loops give rise to orbital-like magnetic moments within $\rm CuO_2$ plaquettes. It is worth noticing that the staggered magnetization  also couples to the uniform magnetization, yielding the weak singularity in the uniform susceptibility at the ordering temperature \cite{Gronsleth}, observed experimentally \cite{Leridon}. The overall symmetry of the magnetic neutron scattering pattern is actually consistent with the  fourfold degenerated CC-$\theta_{II}$ phase of Ref. \cite{Varma}.

Owing to the confinement of the current loops within CuO$_2$ planes, orbital-like magnetic moments are expected to be parallel to the $c$-axis. 
However, the experimental observation  of a substantial planar magnetic component \cite{Fauque,Mook,Li} casts some doubt on the validity of the CC-phase as the origin of the {\bf Q}=0 AF state. Nevertheless, several theoretical studies propose different ways to overcome this difficulty. For ${\rm YBa_2Cu_3O_{6+x}}$, for instance, a spin-orbit coupling effect can explain the appearance of a weak planar magnetic component \cite{Aji}. In addition, an extremely weak ferromagnetism may accompany the CC order in magneto-optical measurements (Kerr effect) \cite{Kapitulnik}. For a ${\rm CuO_6}$ octahedron (realized in the ${\rm HgBa_2CuO_{4+\delta}}$ system), variational Monte-Carlo numerical simulations show that circulating currents could be delocalised over all oxygen orbitals, even the apical ones \cite{Weber}. In the CC-$\theta_{II}$ phase, there are 4 states corresponding to different configurations of the current loops within the CuO$_2$ plaquettes. In the initial CC-theory, this leads to 4 classical domains. Once excitations are taken into account\cite{Varma-INS}, it turns out that the ground state is made of the quantum superposition of the 4 states. Then, these quantum corrections can account for the observation of orbital-like magnetism with a planar magnetic component, as observed in polarized neutron experiments. In addition, the theory predicts specific quasi non-dispersive collective magnetic excitations and one of them has been recently observed in polarized inelastic neutron scattering measurement in ${\rm HgBa_2CuO_{4+\delta}}$ \cite{Li-INS}. Thus, to date, the CC-phase remains the only available theoretical approach able to account for the {\bf Q}=0 AF order observed by polarized neutron diffraction.

While the {\bf Q}=0 AF order appears 3D and is of long-range character in ${\rm YBa_2Cu_3O_{6+x}}$  \cite{Fauque,Mook} and ${\rm HgBa_2CuO_{4+\delta}}$ \cite{Li}, its stability in ${\rm La_{2-x}Sr_{x}CuO_4}$ seems more questionable. In this system, for a hole doping $p$=0.085, the {\bf Q}=0 AF order  remains 2D and is short range \cite{Baledent}. It further sets in below $T_{mag} \sim$ 120 K, a temperature much smaller that the expected value of $T^{\star}$ for such a low hole doping. In this cuprate family, the expansion of the {\bf Q}=0 AF order is likely to be limited by the occurence of a competing electronic instability \cite{Baledent}. In ${\rm La_{2-x}Sr_{x}CuO_4}$, this competing phase could take the form of fluctuating stripes which tend to become static close to a hole doping of p=1/8  or in presence of disorder (impurities, random potential, point defects) and external perturbations such as a uniform magnetic field \cite{Tranquada,Kivelson}. Following numerical simulations \cite{Gabay,Scalapino} indicating that a CC-phase  could develop in two-leg ladders at least at short range, it has been suggested that currents loops could be confined within charge stripes in ${\rm La_{2-x}Sr_{x}CuO_4}$ \cite{Baledent}.

This observation rises questions about the stability of the orbital-like  magnetic order  in strongly underdoped ${\rm YBa_2Cu_3O_{6+x}}$. The order has been observed down to a hole doping of $p$=0.09. At slightly lower hole doping $p_{MI}\sim$0.085, a metal-insulator (MI) transition is inferred from quasiparticule heat transport \cite{Sun}: this critical doping is assumed to be related to  the freezing of stripe correlations at low temperature, as revealed by the in-plane resistivity anisotropy (the blue dashed area in Fig.~\ref{Fig-intro}). In addition, the transport data has been interpreted as a critical divergence of the cyclotron mass ($m^{\star}$), yielding a collapse of the Fermi temperature ($\propto 1/m^{\star}$) around $p_{MI}$ \cite{Sebastian}. However, a different conclusion has been recently drawn from other transport measurements\cite{proust2010} where the metal-insulator crossover at $p_{MI}$ is claimed to be driven by a Lifshitz transition, namely it would correspond to a change in Fermi-surface topology at the critical concentration $p_{MI}$ where the electron Fermi pockets vanish. Without going into that specific debate, it is here important  to stress that all transport measurements point towards a critical doping, $p_{MI}$, with different electronic properties on both sides. At the same doping level, it is worth to remind that the spin excitation spectrum also changes drastically \cite{Rossat,Dai,Y645,Y66}. Likewise, the spontaneous appearance of an a-b anisotropy in spin correlations provides strong evidences of electronic liquid crystal (ELC) phases near $p_{MI}$ \cite{Y645,Y645-Field,Y66-Zn,ELC}. Under the generic name of ELC phases, one usually group all states of correlated quantum electronic systems that break spontaneously either rotational invariance or translation invariance \cite{Fradkin}. Using such a terminology, a fluctuating stripes state breaking the C4 rotational invariance would be classified as a nematic ELC phase, whereas a static stripes state that further break the translation invariance would correspond to a smectic ELC state.

We here report a study of the stability of the 3D {\bf Q}=0 AF order in ${\rm YBa_2Cu_3O_{6+x}}$ for two distinct cases: (i) at low hole doping for $p$=0.08 ($p < p_{MI}$), (ii) at larger hole doping $p$=0.12 ($p > p_{MI}$) in the presence of a disorder introduced through substitution of non magnetic Zn impurities (Fig~\ref{Fig-intro}). Here we report a polarized neutron scattering study of the {\bf Q}=0 AF state in two samples ${\rm YBa_2Cu_3O_{6.45}}$ (Y645, $p$=0.08) and  ${\rm YBa_2(Cu_{0.98}Zn_{0.02})_3O_{6.6}}$ (Y66-Zn, $p$=0.12), where the fingerprints of an ELC phase has already been obtained in previous unpolarized inelastic neutron scattering experiments \cite{Y645,Y66-Zn}.

\section{\label{Exp} Experimental details}
\subsubsection*{\label{set-up} Sample and experimental set-up}

\begin{figure}[t]
\includegraphics[width=8cm,angle=0]{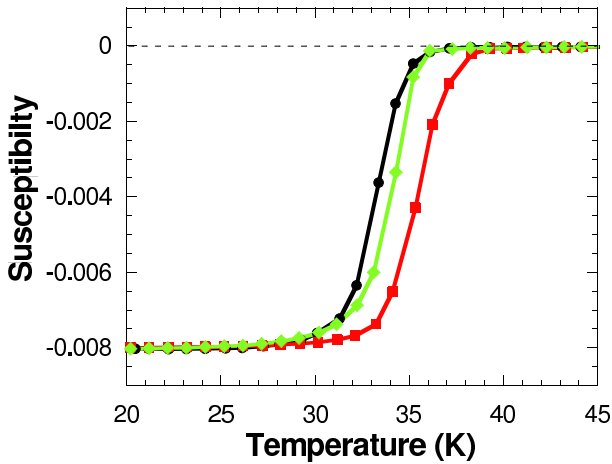}
\caption {
(color online) SQUID susceptibility from 3 different representative single crystals constituting our sample array of ${\rm YBa_2Cu_3O_{6.45}}$ ($p$=0.08, $T_c$=35 K).
}
\label{Fig-SQUID}
\end{figure}

The samples are arrays of co-aligned tiny single crystals glued on a Silicon plate (Y645) or on an Al-grid (Y66-Zn). They are detwinned, with a detwinning ration of  1/10 for Y645 \cite{Y645} and 1/4 for Y66-Zn \cite{Y66-Zn}. Both samples consist of $\sim$ 100 co-aligned and detwinned single crystals. For the Y645 sample array, the sample preparation has been previously detailled in the supporting Online Material of ref. \cite{Y645}. $T_c$ was measured using SQUID magnetometry in a large fraction of the individual single crystals, yielding the average values of $T_c$=35 K (Y645) \cite{Y645} and $T_c$=30 K (Y66-Zn) \cite{Y66-Zn}. Fig. \ref{Fig-SQUID} shows SQUID measurements or three representative single crystals. All crystals were characterized by magnetometry and were found to exhibit superconducting transition temperatures $T_c$ = 35 K with transition widths of 2--3 K, testifying to their high quality. The  bulk $T_c$ of the mosaic sample was further crosschecked using the neutron spin depolarization technique \cite{Fong} on the (004) Bragg reflection. The lattice parameters, a = 3.8388 \AA, b = 3.8747 \AA, and c = 11.761 \AA, correspond to the chemical composition ${\rm YBa_2Cu_3O_{6.45}}$, from which a hole concentration of 0.08 $\pm$ 0.005 per CuO$_2$ plane is deduced using Ref.~\cite{Doping-YBCO} and the $T_c$ transition widths (Fig. \ref{Fig-SQUID}). Both samples exhibit thus a similar $T_c$. For Y66-Zn, the substitution of 2\% Zn/Cu reduces $T_c$ from 64 K in pure Y66 \cite{Fauque} down to 30 K in Y66-Zn, while keeping the hole doping unchanged (p$\simeq$ 0.12). For Y645, the small $T_c$ value of 35 K is due to the reduction of the number of doped hole per $\rm CuO_2$ planes (p$\simeq$ 0.08). 

Polarized neutron scattering measurements were performed on the triple-axis spectrometer 4F1 at the reactor Orph\'ee in Saclay (France). The polarized neutron scattering set-up is similar to the one used in previous experiments on the same topic \cite{Fauque,Mook,Li}: the incident neutron beam is polarized using a polarizing super-mirror and the polarization of the scattered beam is analysed using a Heusler analyzer. Standard XYZ-Helmholtz coils guide the neutron spin polarisation on the sample. The experimental set-up further includes, on the incoming neutron beam, a Mezei flipper for flipping the neutron spin direction, and a pyrolytic graphite filter for eliminating higher harmonics. For the polarized diffraction measurements, the incident and final neutron wave vectors are set to 2.57~\AA$^{-1}$. Samples were attached to the cold head of a 4K-closed cycle refrigerator and aligned in the [010]/[001] scattering plane, so that tranferred wave vectors $\bf{Q}$ of the form (0,K,L) are accessible. It is important to notice that, at variance with studies carried with twinned samples, the [100] and [010] directions are not equivalent anymore in the case of the present study on detwinned samples.

\subsubsection*{\label{polar} Polarized neutron measurements and data analysis}

The magnetic signal we are looking for is induced by a {\bf Q}=0 AF order, implying that the magnetic scattered intensity is superimposed onto the nuclear scattering on Bragg reflections. The magnetic intensity is typically 4 orders of magnitude weaker than the strongest nuclear Bragg reflections. Therefore, polarized neutron technique is essential to disentangle nuclear and magnetic contributions on Bragg reflections. Owing to the weakness of the magnetic intensity, the success of this measurement relies on the quality of polarization of the neutron beam, given by the flipping ratio ($R_0$).
To this end, the scattered intensity on a selected Bragg reflection is systematically measured in the spin-flip (SF) and non-spin-flip (NSF) channels, with 3 orthogonal neutron spin polarizations: Hx,Hy,Hz. For Hx and Hy, the neutron spin polarization is respectively parallel and perpendicular to $\bf{Q}$ in the scattering plane. For Hz, the neutron spin polarization is perpendicular to the scattering plane. In the rest of the manuscript, the indices SF,NSF and $\alpha$={x,y,z} indicate to which channel and to what kind of polarizations the measurements correspond.

The magnetic intensity, $I_{mag}$ and the inverse flipping ratio $R^{-1}_{\alpha}(T)$ are given by the following equation:
\begin{equation}
R^{-1}_{\alpha}(T)=\frac{I_{SF,\alpha}}{I_{NSF,\alpha}}= \frac{I_{mag,\alpha}}{I_{NSF,\alpha}}+ R^{-1}_{0,\alpha}(T)
\label{Imag}
\end{equation}
$I$ stands for the Bragg intensity and $R^{-1}_{0,\alpha}(T)$ is the inverse flipping ratio in the absence of any magnetic signal.

The study of the magnetic signal is ususally performed on Bragg reflections (0,1,L) and (1,0,L) with integer L values. For twinned samples, (0,1,L) and (1,0,L) nuclear Bragg reflections are equivalent. At variance, using a detwinned sample, the nuclear scattering at (0,1,L) is significantly weaker than at (1,0,L), since the nuclear response associated with CuO chains is absent. In the SF channel, the polarization leakage from the NSF channel is therefore substentially reduced for (0,1,L), allowing a better detection of the magnetic scattering.

According to theory, $R^{-1}_{0,\alpha}$ should not depend on $\bf{Q}$, neutron polarization and temperature. In practice, that is not the case due the imperfect experimental setup and a shift of the sample position within the guide field when changing temperature. Thus, for each Bragg reflection (0,1,L) and  for a given polarization $\alpha$ , one calibrates the magnitude of $R^{-1}_{0,\alpha}$ at high temperature where the magnetic signal is absent. The temperature dependence of $R^{-1}_{0,\alpha}$ is measured at the larger wave vector $\bf{Q}$= (0,2,0) or (0,2,1) where the magnetic scattering is vanishingly small.

In the SF channel, the scattered intensity is obviously dominated by this polarization leakage at all temperatures. Nevertheless, one may also need to pay attention for the background on top of which the Bragg scattering develops. Indeed, the temperature dependence of this background can blur the extraction of the temperature dependence of the magnetic intensity. For each studied Bragg reflection (0,1,L), a systematic measurement of the background is therefore performed at (0,0.9,L) away from the Bragg scattering in both the SF and NSF channels and substracted from the scattered intensity measured at the Bragg reflection. This background substraction was not carried out in previous studies \cite{Fauque,Mook}: it actually corresponds to a refinement of the method used to extract the magnetic signal. It allows a better determination of the temperature dependence of the magnetic signal, especially when this signal is weaker than in the previous studies: this is precisely the case in the present study.

Once the magnetic intensity is determined, it can be calibrated in absolute units (mbarns) by comparison with the intensity of the (0,0,4) nuclear Bragg reflection (7 barns): this intensity is indeed hardly dependent on the oxygen content and provides a good reference for calibration. This calibration procedure is the same as the one used in previous studies \cite{Fauque,Mook}.

\section{\label{results} Experimental results}

\subsubsection*{{\bf Q}=0 AF order at low hole doping}

To begin with, let us consider the hole doping effect on the stability of the 3D {\bf Q}=0 AF order. Previous studies established that $T_{mag}$ and the value of the order magnetic moment at low temperature grew continuously in ${\rm YBa_2Cu_3O_{6+x}}$,  when reducing the oxygen content from x=0.75 down to x=0.5 \cite{Fauque}. In Y65 ($T_c$=54 K), the hole doping is p$\sim$0.09 ($> p_{MI}$) and $T_{mag}$ is close to room temperature \cite{Fauque}. The study of our Y645 sample allows us to track the evolution of the magnetic signal at lower oxygen content (x=0.45) and slightly lower hole doping ($p$=0.08).

\begin{figure}[t]
\includegraphics[width=8cm,angle=0]{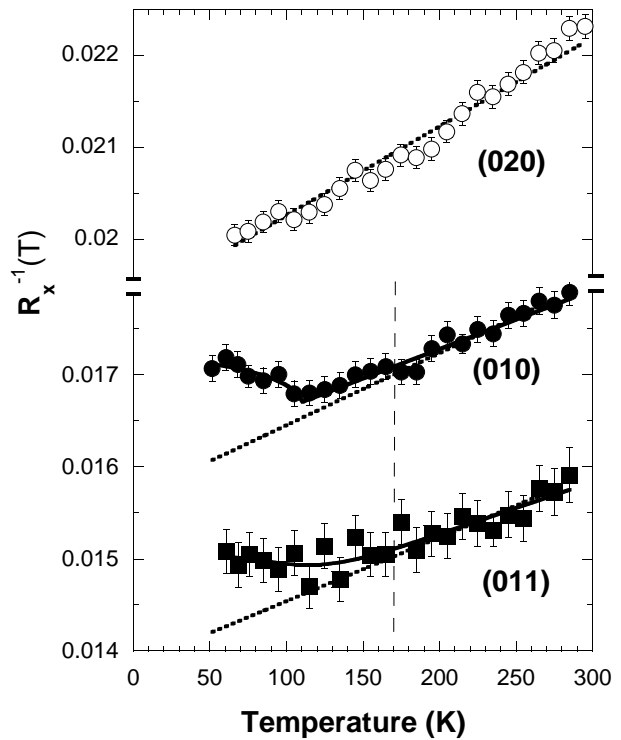}
\caption { $\rm YBa_2Cu_3O_{6.45}$ - Temperature dependence of $R^{-1}_{x}$ at different Bragg reflections.
Solid lines are guides to the eye. The dotted lines indicate $R^{-1}_{0,x} \propto (1+A \times T)$ with $A=0.0005 K^{-1}$.
}
\label{Fig4}
\end{figure}

Figure ~\ref{Fig4} shows  $R^{-1}_{x}(T)$ at the Bragg reflections (0,2,0), (0,1,0) and (0,1,1). At (0,2,0), $R^{-1}_{x}(T)$ exhibits an almost linear decrease on cooling down. This allows us to set the T-dependence of the reference inverse flipping ratio: $R^{-1}_{0,x} \propto (1+A \times T)$ with $A=5x10^{-4} K^{-1}$. At variance,  at (0,1,0) and (0,1,1) , $R^{-1}_{x}(T)$ increases smoothly on top of $R^{-1}_{0,x}(T)$ below $T_{mag} \simeq$ 170 K $\pm$ 30 K. According to Eq.~\ref{Imag}, this indicates the gradual appearance of the magnetic intensity below $T_{mag}$. Figure~\ref{Fig4} highlights two important features. For a small reduction of the hole doping, $T_{mag}$ steeply drops from 300 K ($p > p_{MI}$) down to 170 $\pm$ 30K ($p < p_{MI}$). Meanwhile, the magnetic intensity at (0,1,0) and (0,1,1) decreases from 9 $\pm$1.8 mbarns and 2.9$\pm$0.3 mbarns in Y65 \cite{Fauque} down to 1.7$\pm$0.2 mbarns and 0.3$\pm$0.2 mbarns here in Y645. Furthermore, the T-dependence of the magnetic intensity suggests that $T_{mag}$ should be viewed as an onset temperature rather than a net transition temperature. Our data therefore demonstrate that the 3D {\bf Q}=0 AF order exhibits a sudden decrease in strongly underdoped Y645, i.e slightly below $p_{MI}$, and vanishes at lower doping on approaching the insulating state. 

It is generally believed that $T^{\star}$ grows continuouly when decreasing the hole doping and should be much larger than room temperature upon approaching the Mott AF phase. In this case, our study shows that $T_{mag}$ does not seem to match $T^{\star}$ anymore at low doping. However, the exact doping behavior of $T^{\star}$ at such low doping might have been overlooked. It is indeed worth pointing out that the $T^{\star}$ determination in strongly underdoped ${\rm YBa_2Cu_3O_{6+x}}$ is not very accurate neither in NMR Knight shift \cite{Alloul-1989} nor with resistivity measurements \cite{Ito}. In Ref.~\cite{Ito}, samples with the nominal oxygen content x=0.45 exhibit much higher $T_c$ than our Y645, suggesting that their hole doping could be $p\sim$0.1 rather than $p$=0.08. Furthermore, measurements above room temperature are almost impossible owing to the onset of oxygen mobility in the samples. For resistivity measurements carried out on  strongly underdoped  samples \cite{Ando}, the detection of an anomaly around our value of $T_{mag}$ could be hidden by the upturn of the resistivity at low temperature which characterized these samples. Therefore, what we found here for $T_{mag}$ might be the actual doping behavior for $T^{\star}$.

For a doping level of $p$=0.085, the {\bf Q}=0 AF order develops at a low temperature in the monolayer system $\rm La_{2-x}Sr_xCuO_4$ ($T_{mag}$=120 K)\cite{Baledent}, while in bilayer system $\rm YBa_2Cu_3O_{6+x}$ the order develops slowly $T_{mag}\sim$170 K and is further enhanced below $\sim$120 K (Fig.~\ref{Fig4}). In the former system, the order is 2D and at short range. In the later system,  the magnetic neutron scattering intensity is still observed on Bragg reflections, suggesting that 3D correlations survive: Attempts to detect a 2D magnetic intensity at $\bf{Q}$=(0,1,L) with non-interger L values do not reveal any signal. In $\rm La_{2-x}Sr_xCuO_4$, the body centered crystal structure, the intrinsic disorder due to Sr/La substitution and an antiparallel stacking of charge stripes along the $c$-axis could contribute to the reduction of the {\bf Q}=0 AF correlations along the $c$-axis.

\subsubsection*{Dilution of the {\bf Q}=0 AF order through substitution of quantum impurities}

 We study in this section the effect of non-magnetic Zn impurities on the {\bf Q}=0 AF order for a doping level above $p_{MI}$. While $T_c$ is strongly reduced by non magnetic impurity, neither hole doping nor $T^{\star}$ are modified. However spin correlations are known to be enhanced locally around Zn impurities, according to local probe measurements \cite{Mahajan,Bobroff}. The local spin correlations are expected to compete with the {\bf Q}=0 AF order.

\begin{figure}[t]
\includegraphics[width=8.5cm,angle=0]{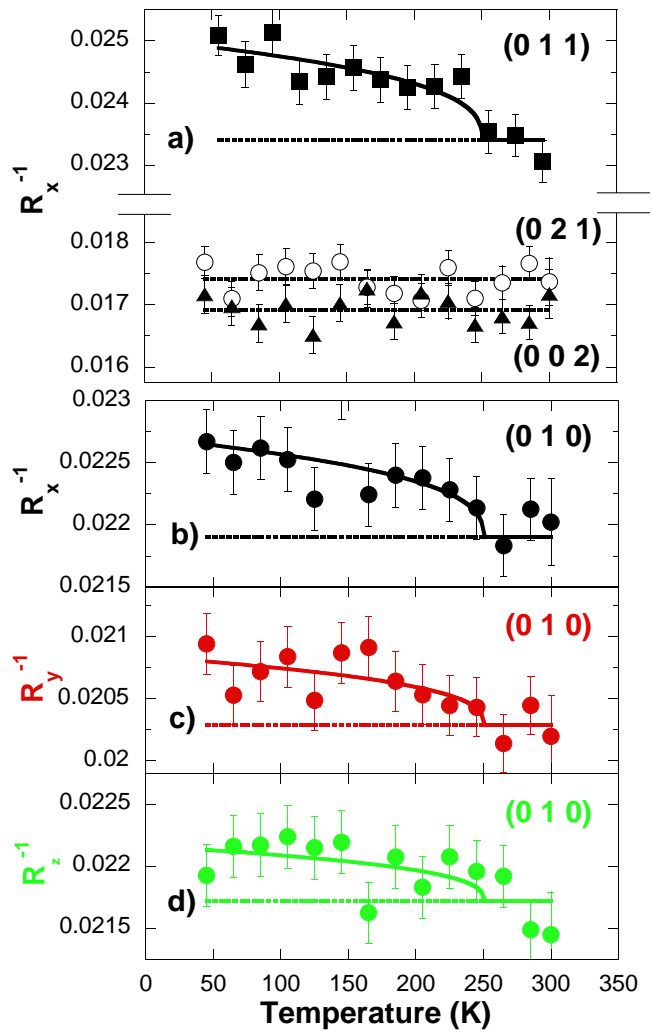}
\caption {
(color online) ${\rm YBa_2(Cu_{0.98}Zn_{0.02})_3O_{6.6}}$ - a) Temperature dependence of $R^{-1}_{x}$ at different Bragg reflections. The dotted lines indicate $R^{-1}_{0,x} (T)$ which is a constant for that set of measurements. b-d) Temperature dependence of $R^{-1}_{\alpha}$ at (0,1,0) for different polarizations: b) Hx (black), c) Hy(red), d) Hz(green). The solid lines indicate the appearance of the magnetic signal at $T_{mag} \sim $ 250 K characterized by the temperature dependence $\propto (1- T/T_{mag})^{2 \beta}$ with $\beta$=0.18 \cite{Mook}. For measurements on the Bragg reflections (0,2,1) and (0,0,2), the background is measured at ${\bf Q}$=(0,2$\pm$0.1,1) and (0,0,2$\pm$0.2), respectively.
}
\label{Fig2}
\end{figure}

As in Y66 \cite{Mook}, the T-dependence of the reference inverse flipping ratio, $R^{-1}_{0,x}(T)$, is measured on the Bragg reflection (0,2,1). Fig.~\ref{Fig2}.a shows that $R_{0,x}^{-1}(T)$ is actually a constant for that particular experiment. Furthermore, $R_{x}^{-1}(T)$ is also temperature independent at ${\bf Q}$=(0,0,2). This is a characteristic feature of the {\bf Q}=0 AF order, which does not yield any magnetic scattering on Bragg reflections (0,0,L) with integer L values \cite{Fauque,Baledent}. At variance, $R_{0,x}^{-1}(T)$  starts growing below $T_{mag}$=250$\pm 20$K at ${\bf Q}$=(0,1,1) (Fig.~\ref{Fig2}.a) and (0,1,0)(Fig.~\ref{Fig2}.b).  At (0,1,0), one can check that the polarization sum rule $R^{-1}_{x}=R^{-1}_{y}+ R^{-1}_{z}$ (valid for a magnetic signal only in absence of any chirality) is fulfilled within error bars: This demonstrates of the magnetic nature of the signal that develops below $T_{mag}$.

In Y66-Zn, we observe a magnetic signal reminiscent of the one reported in pure Y66 \cite{Fauque,Mook}. It displays the same kind of structure factor, characterized by the appearance of a magnetic response on (0,1,L) Bragg reflections and the absence of a magnetic response at on Bragg reflections (0,0,L). The polarization analysis of the data is consistent with magnetic moments oriented at $\sim$45$^{o}$ with respect to the $c$-axis, in agreement with similar analysis in the Zn free coumpound \cite{Fauque,Mook}. The temperature dependence can be described using the a power law $(1-T/T_{mag})^{2\beta}$ (with $\beta=0.185 \pm 0.06$) as reported for pure Y66 (Fig.~\ref{Y645-Tdep}). Finally, the magnetic order sets in at almost the same temperature $T_{mag}$ as in pure Y66 \cite{Fauque,Mook}. This is in agreement with Knight NMR measurements, which demonstrated that the onset pseudo-gap temperature T$^{\star}$ was not affected by substitution of non magnetic Zn impurities \cite{Alloul-Zn}.

\begin{figure}[t]
\includegraphics[width=6cm,angle=0]{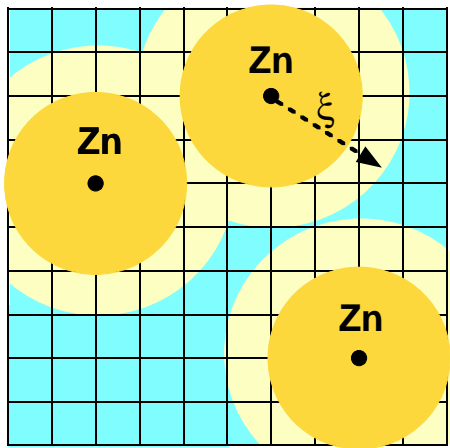}
\caption {
(color online) Array of 10$\times$10 CuO$_2$ plaquettes. The grid indicate the plaquettes where the {\bf Q}=0 AF order should still be present. The black full dots stand for the Zn impurities (3\%  per Cu in CuO$_2$ planes) which are surrounded by an area where the {\bf Q}=0 AF order is supposed to vanish (large orange disks). In addition, quasi-1D incomensurate spin fluctuations are restored at low energy around Zn impurities: their characteristic correlation length $\xi$ is about 2.9 $a$ \cite{Y66-Zn}.
The yellow circles around impurities suggest a possible overlap between static {\bf Q}=0 AF correlations and dynamical incomensurate spin correlations.
}
\label{Fig3}
\end{figure}

Nevertheless, dilute quantum impurities induce a net perturbation of the {\bf Q}=0 AF order, which becomes apparent after a more quantitative comparison of data in Y66 \cite{Mook} and Y66-Zn (Fig.~\ref{Y645-Tdep}). The magnetic intensities at low temperature on the (0,1,0) and (0,1,1) Bragg reflections are 5.4$\pm$0.4 mbarns and 1.4$\pm$0.2 mbarns respectively in pure Y66. They drop down to 2$\pm$0.5 mbarns and 0.9$\pm$0.2 mbarns under  Zn substitution. Since the neutron scattering measurement is a bulk measurement, the reduction of the magnetic intensity can be accounted for by a reduction of the sample fraction occupied by the magnetic order. In a dilution model, also called "swiss cheese model" \cite{Alloul-review}, the magnetic order within $\rm CuO_2$ planes is preserved far from Zn impurities: Figure \ref{Fig3} gives a cartoon picture of the inhomogenous magnetic distribution induced by the development of protected clusters around local defects. Far from Zn impurities, the ordering temperature and the magnitude of the order moment remain those of the pure system: they are exclusively controled by the hole doping, which is not modified by Zn subtitution. At variance, the magnetic order locally vanishes close to Zn impurities. Considering , on the one hand, an overall reduction of the magnetic intensity by a factor of 0.5$\pm$0.2 and , one the other hand, that the magnetic intensity in neutron diffraction measurements is proportional to the square of the ordered magnetic moment (M), one can estimate that the magnetic order is present in 70$\pm$10\% of the $\rm CuO_2$ plaquettes. In a dilution model, the magnetic moment reads:
\begin{equation}
M(z)=M(0) (1-\gamma z)
\end{equation}
with $z$ the number of impurities per Cu in $\rm CuO_2$ planes: in $\rm YBa_2(Cu_{1-y}Zn_y)_3O_{6+x}$, $z=\frac{3}{2} y$, i.e 3\% in our case. $\gamma$ stands for the efficiency coefficient which defines the spatial extension of the disorder introduced around a given impurity. We obtain a value $\gamma$=10$\pm$3, implying  that a Zn induced disorder spreads up to the third nearest Cu neighbors around the impurity site (Fig.~\ref{Fig3}), in perfect agreement with previous estimation from NMR \cite{Mahajan,Bobroff} and inelastic neutron scattering \cite{Y66-Zn}.

\section{\label{discussion} Discussion}

\subsubsection*{Hole doping induced decay of the magnetic order versus dilution effect}

\begin{figure}[t]
\includegraphics[width=8cm,angle=0]{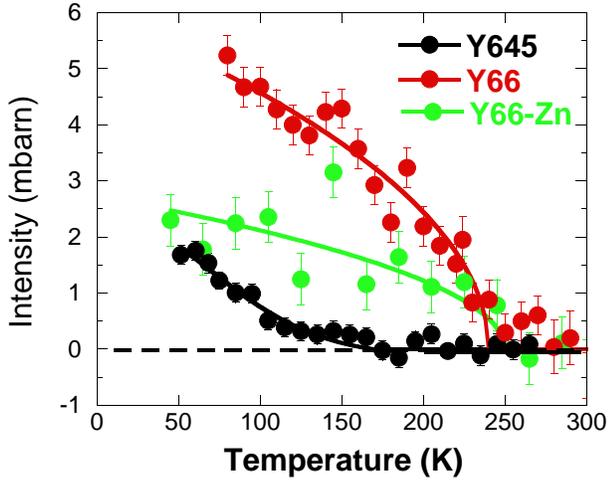}
\caption { (color online) Temperature dependence of the magnetic intensity $I_{mag}$ at ${\bf Q}$=(0,1,0) (or (1,0,0) for twinned samples): $\rm YBa_2Cu_3O_{6.45}$ (black - Y645, $T_{mag}\sim$ 170 K ), $\rm YBa_2Cu_3O_{6.6}$ (red - Y66, $T_{mag}\sim$ 235 K \cite{Mook}), $\rm YBa_2(Cu_{0.98}Zn_{0.02})_3O_{6.6}$ (black - Y645, $T_{mag}\sim$ 250 K).    
}
\label{Y645-Tdep}
\end{figure}

Fig.~\ref{Y645-Tdep} reports the temperature dependence of the magnetic intensity at (0,1,0) calibrated in absolute units in pure Y66, Y66-Zn and Y645: it summarizes our main observations. In Y66-Zn, $T_{mag}$ is not affected by Zn substitution and remains close to the value reported in pure Y66. The net reduction of the low temperature ordered magnetic moment can be ascribed to a reduction of effective sample fraction occupied by the {\bf Q}=0 AF order owing to the local destruction of {\bf Q}=0 AF correlation in the vicinity of quantum impurities. In this swiss cheese scenario, $T_{mag}$ and the ordered magnetic moments are preserved far from Zn impurity, being essentially controlled by the hole doping level which is not modified by Zn substitution. In Y645, the low temperature order moment is of the same magnitude as in Y66-Zn, but $T_{mag}$ is strongly reduced. $I_{mag} (T)$ does not follow the power law $(1-T/T_{mag})^{2\beta}$ (with $\beta=0.185 \pm 0.06$) as observed at larger doping in Y66 \cite{Mook} (Fig.~\ref{Y645-Tdep}). $I_{mag}$ increases more gradually on cooling down (Fig.~\ref{Y645-Tdep}). This suggests that there could be a distribution of ordering temperatures or that the {\bf Q}=0 AF correlation length still increases below $T_{mag}$. In general, for magnetic systems, the low temperature magnetic moment $M$($\propto \sqrt{I_{mag}}$ in neutron diffraction) \cite{squires} and the ordering temperature are expected be proportional.
From Y65 to Y645, only $\sim$20\% of the magnetic intensity remains at low temperature, corresponding to a reduction of the low temperature ordered moment by a factor $\sim$ 2.3. A scaling relationship $M\propto T_{mag}$ would implies a similar decrease of $T_{mag}$ from 300 K (Y65) down to 130 K (Y645). This value is consistent with the observed $T_{mag} \simeq$ 170 $\pm$ 30 K in Y645. One can therefore conclude that the reduction of both $T_{mag}$ and the low temperature ordered moment can be ascribed to a uniform weakening of the 3D {\bf Q}=0 AF order below $p_{MI}$=0.08.

\subsubsection*{Competition with electronic liquid crystal phases}

In strongly underdoped ${\rm La_{2-x}Sr_xCuO_4}$ ($p$=0.085), a 2D short range {\bf Q}=0 AF order develops below $T_{mag}\sim$ 120 K \cite{Baledent}. In addition, the evolution in the same temperature range of the incommensurate (IC) spin fluctuations around the planar AF wave vector ${\bf Q}_{AF}$=(0.5,0.5) is of particular interest. In twinned $\rm La_{2-x}Sr_xCuO_4$, these fluctuations can be observed at planar wave vectors ${\bf Q}_{IC}$= ${\bf Q}_{AF}$ $\pm$($\delta$,0)=${\bf Q}_{AF}$ $\pm$(0,$\delta$). They are usually interpreted as the fingerprint of fluctuation stripes. Surprizingly, the intensity of these IC fluctuations displays an upturn below $T_{mag}$ and the IC parameter $\delta$ starts increasing below $T_{mag}$ like an order parameter. This study suggests that the vicinity of a spin and/or charge density wave state is the main cause of the  limitation of {\bf Q}=0 AF correlations.

\begin{figure}[t]
\includegraphics[width=8.5cm,angle=0]{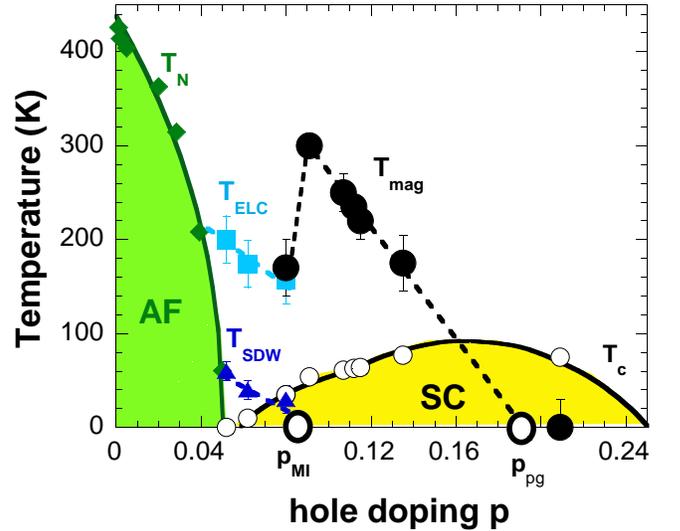}
\caption {
(color online) ${\rm YBa_2Cu_3O_{6+x}}$ - Magnetic phase diagram as a function of hole doping ($p$) (see Fig~\ref{Fig-intro}). At low doping, when the system exhibits an insulating-like behavior at very low temperature, a nematic electronic liquid state shows up below $T_{ELC}$ (ligth blue squares) and transforms into the spin density wave (SDW) state below $T_{SDW}$ (dark blue triangles) \cite{ELC}. In the metallic state, the {\bf Q}=0 AF order appears below $T_{mag}$ (black bullets) \cite{Fauque,Mook}. Dashed lines the linear extrapolations of $T_{mag}$ and $T_{SDW}$. The SDW may vanish around critical doping $p_{MI}\sim$0.085. The {\bf Q}=0 AF disappears at larger hole doping close to $p_{pg}\sim$0.19, which is also the end point of the pseudo-gap phase according to thermodynamic measurements \cite{Tallon}. The figure further shows that $T_{mag}$ steeply decreases below $T_{MI}$.
}
\label{Fig5}
\end{figure}

Following this study, we can try to interpret the evolution of the {\bf Q}=0 AF order in Y645 in the light of the spin correlations around ${\bf Q}_{AF}$ present in this material and measured on the same sample \cite{Y645}. In Y645, the low energy spin fluctuation spectrum is characterized by  IC spin fluctuations that spontaneously develop a net a-b anisotropy below $T_{ELC} \sim$ 150 K, at the planar wave vector ${\bf Q}_{IC}$= ${\bf Q}_{AF}$ $\pm$($\delta$,0) ($\delta$=0.045).
At low temperature, these  fluctuations freeze gradually, yielding a quasi-elastic magnetic signal. At the time scale of neutron scattering experiment, a quasi-1D short range spin density (SDW) order is detected below $T_{SDW}\sim$ 30 K, but magnetic correlations become really static below a few Kelvin according to $\mu$SR measurements \cite{Y645}. $T_{ELC}$ is interpreted as the fingerprint of a nematic ELC phase, where the C4 rotational symmetry is spontaneously broken owing to strong electronic correlations \cite{Fradkin}. At low temperature, the lattice translation is further broken in the SDW phase, that can be viewed as (glassy) smectic ELC phase. The complete study of spin correlations in strongly underdoped $\rm YBa_2Cu_3O_{6+x}$ from x=0.3 to 0.45 \cite{ELC}, reveals a continuous decrease of $T_{ELC}$ and $T_{SDW}$ when increasing hole doping (Fig.~\ref{Fig5}). The SDW phase should vanish when approaching  the metal-insulator critical doping $p_{MI} \sim$0.085, which seems to be the quantum critical point associated with the SDW state. Further evidence for quantum criticality is derived from the scaling properties of the dynamical spin correlations of Y645 \cite{Y645-scaling}. In line with this scenario, the scattering function at T = 0 exhibits a spin gap for $p > p_{MI}$ \cite{Y66}, but a dilute concentration of nonmagnetic Zn impurities locally restores the SDW in Y66-Zn \cite{Y66-Zn}.

Figure~\ref{Fig5} shows the evolution of the different temperatures $T_{mag}$, $T_{ELC}$ and $T_{SDW}$ as a function of hole doping for ${\rm YBa_2Cu_3O_{6+x}}$. When decreasing the hole doping in pure ${\rm YBa_2Cu_3O_{6+x}}$, $T_{mag}$ decreases steeply across $p_{MI}$. This indicates that the {\bf Q}=0 order becomes destabilized by the ELC phase below $p_{MI}$.
In Y645 ($p < p_{MI}$), the {\bf Q}=0 AF order and the nematic ELC phase develop in the same temperature range: $T_{mag}\simeq$ 170 K and  $T_{ELC}\sim$150 K. The appearance of static {\bf Q}=0 AF correlation and the changes in IC spin fluctuations almost at the same temperature in both ${\rm YBa_2CuO_{6+x}}$ and ${\rm La_{2-x}Sr_xCuO_4}$ around $p\sim$0.08 suggest the existence of a generic interplay between the {\bf Q}=0 AF order and the electronic instability generating the incommensurate spin correlations: it is conceivable that both phases coexist at low doping. In Y66 ($p > p_{MI}$), the 3D {\bf Q}=0 AF order is fully developped, but the ELC instability is still latent. As a result, a phase separation takes place in Y66-Zn when introducing local defects in CuO$_2$ planes: the {\bf Q}=0 AF order primarily vanishes around defects, while the SDW immediately nucleates around the Zn impurities (Fig.~\ref{Fig3}). Far from Zn impurities, the {\bf Q}=0 AF order remains essentially the same as in the Zn free system. Its ordering temperature and ordered magnetic moments are likely to be controled by the hole doping which remains unchanged under Zn substitution. 

At this stage, it is too early to give a definitive picture of the mixed phase that could develop below $p_{MI}$. Indeed, there is no clear consensus either about the origin of the {\bf Q}=0 AF order, or about the origin of the quasi-1D IC-SDW phase. Up to now, the only model that could account for the {\bf Q}=0 AF state is the circulating current model \cite{Varma} with orbital-like magnetic moment expected primarily along the ${\bf c}$-axis \cite{Varma-INS}. Concerning the SDW instability, one can notice that the evolutions of the incommensurate parameter $\delta$ as a function of hole doping  are different in the monolayer system ${\rm La_{2-x}Sr_xCuO_4}$ and the bilayer system ${\rm YBa_2Cu_3O_{6+x}}$ \cite{ELC}. Such a difference is not expected within a stripe model, but could be accounted for by the spiral-SDW model proposed by O.P. Sushkov \cite{Sushkov}. The orbital-like magnetic moments of a CC-phase and the Cu spins of planar spiral-SDW phase could mix owing to the spin-orbit coupling introduced in Ref.~\cite{Aji}. As a result, Cu spins and orbital-like moments could be further tilted. In this scenario, a tilted spiral-SDW state would be expected. In addition, one cannot exclude a spatial modulation of the order parameter of the CC-phase. This scenario is still highly speculative and further theoretical and experimental works are needed to get a better understanding of such a mixed phase.

\section{\label{conclusion} Conclusion}

We report a polarized neutron diffraction study of the evolution of the 3D {\bf Q}=0 AF order in strongly underdoped ${\rm YBa_2Cu_3O_{6.45}}$ and underdoped ${\rm YBa_2(Cu_{0.98}Zn_{0.02})_3O_{6+x}}$. Our data indicate a dilution of the 3D {\bf Q}=0 AF  through Zn substitution, whereas, the {\bf Q}=0 AF order undergoes a steep drop down when lowering the hole doping below a critical hole doping level $p_{MI}$=0.085. Both phenomena suggest a competition with (nematic and smectic) electronic liquid crystal phases, previously observed in the same samples in inelastic and quasi-elastic neutron scattering measurements. While polarized neutron diffraction has highlighted the appearance of stagerred magnetic moments in CuO$_2$ plaquettes, recent STM measurement in ${\rm Bi_2Sr_2CaCu_2O_{8+\delta}}$ \cite{Bi2212} suggest that the electronic nematicity could also find its origin within the CuO$_2$ plaquettes. Combining all the information should lead condensed matter physicists to reconsider the roles of copper and oxygen in the physics of cuprates and the way an effective single band model should be derived from the 3-band Hubbard model to account for the low energy charge and spin properties in these materials.

\paragraph*{Acknowledgments.}We wish to thank  Beno\^it Fauqu\'e, 
Bernhard Keimer and Chandra Varma for discussions on various aspects related to this work.


\end{document}